\documentclass{article}
\usepackage{spconf,amsmath,graphicx,CJKutf8,CJKnumb,multirow,stfloats}

\title{GLAND SEGMENTATION VIA DUAL ENCODERS AND BOUNDARY-ENHANCED ATTENTION}
\name{{Huadeng Wang}$^{\dagger}$$^{\S}$, {Jiejiang Yu}$^{\S}$, {Bingbing Li}$^{\ddagger}$, {Xipeng Pan}$^{\dagger}$$^{\S}$, {Zhenbing Liu}$^{\dagger}$$^{\S}$, {Rushi Lan}$^{\dagger}$$^{\S}$\sthanks{\scriptsize{* Corresponding author.}}, {Xiaonan Luo}$^{\dagger}$$^{\S}$\thanks{\scriptsize{This work was partially supported by the Guangxi Natural Science Foundation (Grant Nos. 2019GXNSFFA245014 and 2020GXNSFBA238014), the National Natural Science Foundation of China (Grant Nos. 62362014, 62172120, and 62002082), and the Innovation Project of GUET Graduate Education (Grant No. 2023YCXS049).}}}
\address{$^{\dagger}$\scalebox{0.9}{Guangxi Key Laboratory of Image and Graphic Intelligent Processing, Guilin, China}\\
$^{\S}$\scalebox{0.9}{School of Computer Science and Information Security, Guilin University of Electronic Technology, Guilin, China}\\
$^{\ddagger}$\scalebox{0.9}{Department of Pathology, Ganzhou Municipal Hospital, Ganzhou, China}}
\begin{document}
\begin{CJK}{UTF8}{gbsn}
%
\maketitle 
\begin{abstract}
Accurate and automated gland segmentation on pathological images can assist pathologists in diagnosing the malignancy of colorectal adenocarcinoma. However, due to various gland shapes, severe deformation of malignant glands, and overlapping adhesions between glands. Gland segmentation has always been very challenging. To address these problems, we propose a DEA model. This model consists of two branches: the backbone encoding and decoding network and the local semantic extraction network. The backbone encoding and decoding network extracts advanced Semantic features, uses the proposed feature decoder to restore feature space information, and then enhances the boundary features of the gland through boundary enhancement attention. The local semantic extraction network uses the pre-trained DeepLabv3+ as a Local semantic-guided encoder to realize the extraction of edge features. Experimental results on two public datasets, GlaS and CRAG, confirm that the performance of our method is better than other gland segmentation methods.
\end{abstract}
\begin{keywords}
Gland Segmentation, Dual Encoder, Local Semantic Guided Encoder, Boundary Enhance Attention
\end{keywords}
\vspace{-1.4em}
\section{Introduction}
\begin{figure}[htbp]
    \centering
    \includegraphics[width =.48\textwidth]{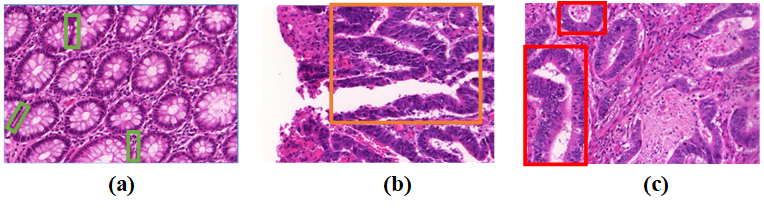}
	\vspace{-2.5em}
    \caption{Morphological structure of glands. (a) The green framed part is the cohesive gland, (b) The orange framed part is the severely deformed malignant gland,  (c) The red framed part is the gland with different shapes.}
	\vspace{-1em}
	\label{sv}
\end{figure}

\vspace{-1em}
\label{sec:intro}
Colorectal adenocarcinoma originates from the glandular tissue of the inner wall of the colon and is a malignant tumor in the digestive system that is often seen in clinical practice[1]. Accuracy of gland instance segmentation in histopathology images is key for pathologists in quantitative analysis of colorectal adenocarcinoma. In the research on gland segmentation based on gland deep learning, pathologists are required to annotate pathological images for training, which requires a lot of time and human resources [2]. Therefore, to reduce the workload of pathologists, It is of great research significance to use computer-aided diagnosis systems to automatically and accurately segment glands from histopathological sections. However, gland segmentation has the following challenges:
(1) There is a great morphological difference between benign glands and malignant glands, and the severe deformation of malignant glands makes it difficult for the network to segment malignant glands (the orange box in Figure \ref{sv}b highlights the severely deformed malignant glands);
(2) Adhesions exist between adjacent glands, making it more difficult to separate each gland from other glands (adhesive glands highlighted by green boxes in Figure \ref{sv}a);
(3) The outline, size, and shape of many glands are inconsistent, and under-segmentation is prone to occur. (different glands highlighted by the red box in Figure \ref{sv}c).       
\vspace{-1em}

Most of the early gland segmentation used some structural knowledge inside the gland, such as morphology-based methods [3] and graph-based methods [4]. These methods only achieved good segmentation results on benign glands. With the great success of deep learning in the field of medical image segmentation, U-Net can effectively capture the multi-scale context information in medical images [5], but for the complex and diverse structures of glands, U-Net does not perform well, Rastogi [6] proposed a colorectal adenocarcinoma histopathological image segmentation algorithm based on U-Net inspired convolutional network, Ding [7] proposed a dual-path colorectal adenocarcinoma pathological image segmentation algorithm, Wen [8] designed Cascade Squeeze Double Attention (CSBA) module to spatial information at different scales, Chen [9] proposed a multi-task learning network based on FCN to simultaneously generate gland regions and contours. Xu [10] A deep three-channel network was proposed to separate dense glands, Graham [11] proposed MILD-Net with both instance and contour segmentation, and Qu [12] proposed a full-resolution network. Although these methods are benign good results have been achieved in gland segmentation, but the gland segmentation of malignant glands and adhesions still fails. The main reason is that these methods have too much downsampling, resulting in the loss of local features at the edge of the gland and the blurring of the gland boundary difficult to identify.

To solve these problems, this paper designs a dual-encoder DEA-Net network based on boundary-enhanced attention for gland segmentation.
The main work and innovations of this paper are as follows:
(1) A dual-encoder DEA-Net is proposed for gland segmentation, in which the pre-trained Deeplabv3+ is used in the local semantic guidance encoder to extract the low-level features of the gland edge, and the main encoder uses a convolutional network to obtain contextual features.
(2) A multi-scale feature fusion method is designed to fuse the features extracted by the dual encoders, which effectively solves the problem of missing glandular edge features in the backbone network.
(3) A feature decoder and a boundary-enhanced attention mechanism are designed on the backbone network to enhance the boundary learning of the gland and better restore the spatial feature information of the gland.
\vspace{-1.5em}
\section{proposed method}
\vspace{-1em}
\label{sec:format}
Our proposed DEA-Net network structure, as depicted in Figure \ref{ff}a, consists of an encoder and a decoder, featuring several crucial modules, including the Local Semantic Encoder (LD), the main encoder, Feature Fusion Module (FFM), the Decoder Feature Block (DFB), Boundary-Enhanced Attention (BEA), and the deep supervision module. We employ two encoders to extract feature information effectively. The Local Semantic-Guided Encoder is dedicated to capturing the edge features of glandular structures, utilizing a pre-trained DeepLabv3+ network that has undergone two downsampling layers. Its role is to compensate for the loss of edge features caused by downsampling in the main network. The backbone encoder consists of 3×3 convolution blocks and max pooling, and the features extracted by both encoders are then fed into the Multiscale Feature Fusion Module (FFM), which enhances the richness of feature context. Subsequently, the Deep Feature Decoder Block (DFB) is employed to retrieve comprehensive feature context information. To ensure the effective restoration of feature information, we have incorporated the Boundary-Enhanced Attention mechanism (BEA) at the end of each decoder to enhance the recovery of glandular boundary information, ultimately achieving more precise glandular segmentation. Finally, the model utilizes bilinear interpolation for upsampling and convolution operations for output. The network employs a variance-constrained cross-entropy loss function to simplify the learning process and executes deep supervision operations on each decoder layer to supervise the recovery of feature information at each level.
\vspace{-1.5em}
\subsection{Feature Fusion Method}
\label{sssec:subsubhead}
\vspace{-0.5em}
Due to significant differences in the sizes of glandular features extracted by two distinct encoders, and recognizing the inherent limitations of a single receptive field in capturing regions of interest, an approach is proposed in this study to enhance the utilization of edge features extracted by locally semantic-guided encoders. To this end, we introduce a multi-scale feature fusion module, denoted as the Feature Fusion Module (FFM) as illustrated in Figure \ref{ff}b, aimed at facilitating improved integration of edge features within the main network architecture. This module comprises multiple scale extractors and a channel attention mechanism. It facilitates the fusion of low-level features \emph{F$_l$} from locally semantic-guided encoders with features \emph{F$_h$} extracted by the main network, resulting in the fused feature map \emph{F$_h^{\prime}$}, Furthermore, a channel attention mechanism is applied to the fused feature map  \emph{F$_h^{\prime}$} to weight it, thereby directing the network's focus towards more salient features. The specific computational process is elucidated as equation \eqref{eq}:
\vspace{-0.5em}
\begin{equation}\label{eq}
{\huge F_a=F_h^{\prime}\scalebox{0.8}{$\otimes$}\sigma_1\left(C_1\left(P_1\left(F_h^{\prime}\right)\right)\right)} 
\end{equation}
In this Equation, \emph{C$_1$} denotes a 1×1 convolution operation, \emph{P$_1$} represents an adaptive global pooling operation, and \emph{$\sigma_1$} signifies the Sigmoid activation function, employed to map the values of the feature map within the range of 0 to 1.
The multi-scale extractor can adaptively extract effective information from multiple receptive fields, thereby enhancing the robustness of the network. The specific operation of this module is shown in equations \eqref{eq2} and \eqref{eq3}:
\begin{figure*}[t]
 	\centering
    \includegraphics[width =0.95\textwidth]{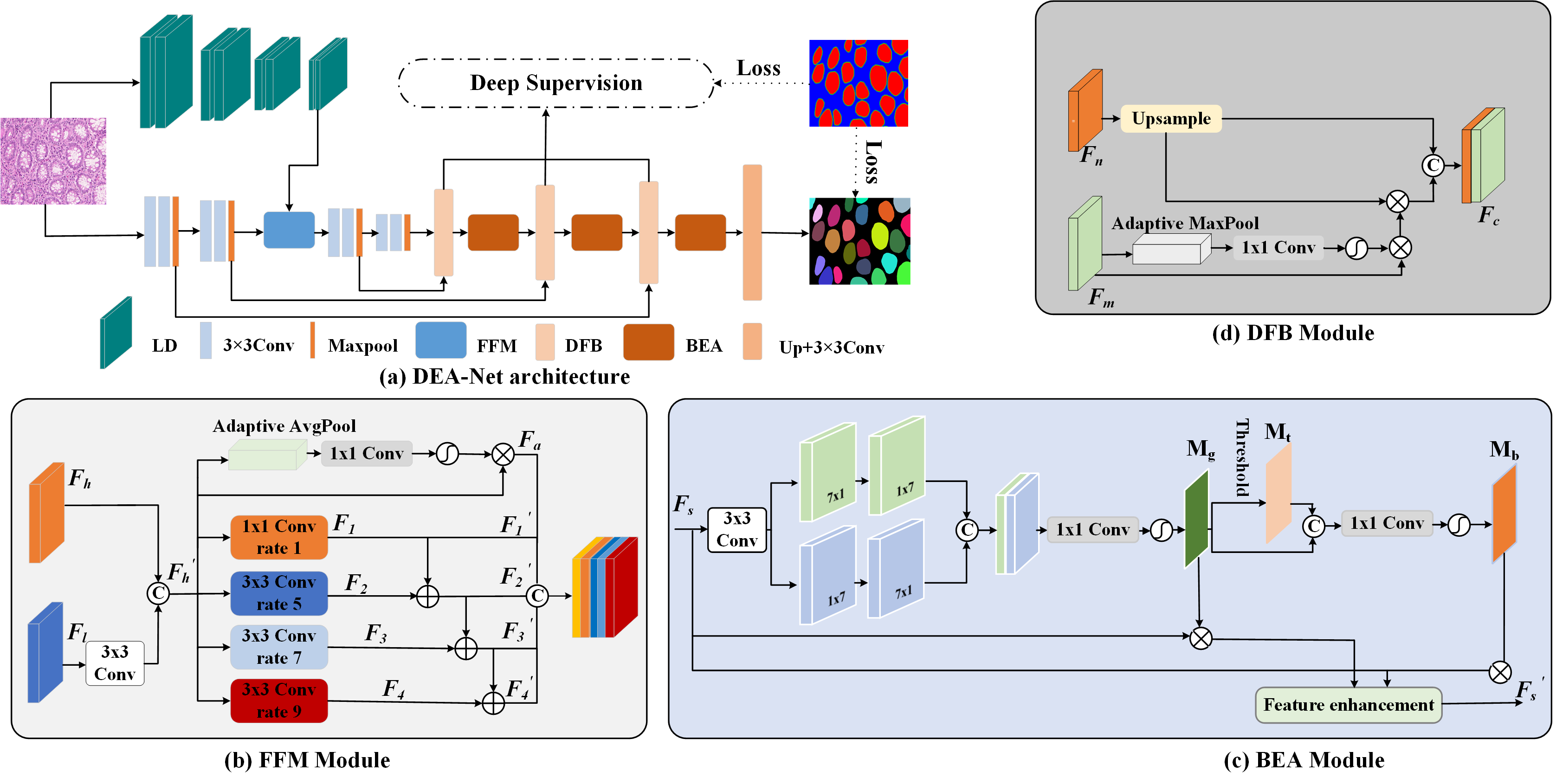}
	\vspace{-1em}
    \caption{DEA-Net Architecture diagram}
	\vspace{-1em}
    \label{ff}
\end{figure*}
\vspace{-0.5em}
\begin{equation}\label{eq2}
{\huge F_i^{\prime}=F_i,i=1}
\end{equation}
\vspace{-1.5em}
\begin{equation}\label{eq3}
{\huge F_i^{\prime}=F_{i-1}\scalebox{0.8}{$\oplus$} F_i,i\in \left [ 2,4 \right ]}
\end{equation}
Where \emph{F$_l$} corresponds to the feature maps obtained from \emph{F$_h^{\prime}$} following convolution operations with diverse dilation rates. Ultimately, a residual feature fusion is conducted on these five distinct-scale feature maps, and the specific procedure is outlined in equation \eqref{eq4}：
\begin{equation}\label{eq4}
 F_m=\sigma_2\left(C_1\left(F_1^{\prime} \scalebox{1.5}{$\text{©}$} F_2^{\prime} \scalebox{1.5}{$\text{©}$} F_3^{\prime} \scalebox{1.5}{$\text{©}$} F_4^{\prime} \scalebox{1.5}{$\text{©}$} F_a\right)\right) \scalebox{1.5}{$\oplus$} G\left(F_h^{\prime}\right)
\end{equation}
In this equation, the symbol \text{©} represents the operation of concatenating feature maps, denoted as \emph{C$_1$} which corresponds to a single 1x1 convolution operation. Additionally, the symbol \emph{$\sigma_2$} denotes the application of the non-linear Rectified Linear Unit (ReLU) activation function.
\vspace{-1em}
\subsection{Deep Feature Decoder Module}
\label{ssec:subhead}
\vspace{-0.5em}
To enhance the restoration of information within the fused dual encoder, we have devised a deep feature decoder (DFB) specifically tailored for the recovery of spatial information about the gland. The detailed design of this decoder is illustrated in Figure \ref{ff}d. DFB is divided into two branches, and the specific process is shown in equation \eqref{eq5} and \eqref{eq6}:
\vspace{-0.5em}
\begin{equation}\label{eq5}
{\huge F_{\mathrm{m}}{ }^{\prime}=F_m \scalebox{0.8}{$\oplus$} \sigma_1\left(C\left(P_2\left(\left(F_m\right)\right)\right)\right)}
\end{equation}
\vspace{-1em}
\begin{equation}\label{eq6}
F_c = \left ( F_{\mathrm{m}}{ }^{\prime}\scalebox{1.5}{$\oplus$} U_p\left ( F_n \right ) \right )\scalebox{1.5}{$\text{©}$}U_p\left ( F_n \right ) 
\end{equation}
Among them, \emph{$F_m$} is the low-level feature map extracted by the encoder, \emph{$P_2$} is the adaptive maximum pooling operation,  \emph{$F_n$} is the high-level feature map with rich information, and \emph{$U_p$} represents the bilinear interpolation upsampling operation.
\vspace{-1em}
\subsection{Boundary Enhanced Attention Mechanism}
\label{sssec:subsubhe}
\vspace{-0.5em}
The boundary recognition of glands plays a key role in the network's accuracy of gland segmentation results. To better enable the network to recognize the boundary features of glands, we design a gland boundary-enhanced attention mechanism behind the decoder DFB (BEA). Through this adaptive threshold attention mechanism, we increase the weight of gland boundary pixels, so that the network is more focused on the boundary of the gland. The specific design of BEA is shown in Figure \ref{ff}c. The specific process is shown in equation \eqref{eq7} and \eqref{eq8}:\vspace{-0.5em}
\begin{equation}\label{eq7}
F_{\mathrm{g}}=C_v\left(C_l\left(C_2\left(F_{\mathrm{s}}\right)\right)\right)\scalebox{1.2}{$\text{©}$}  C_l\left(C_v\left(C_2\left(F_{\mathrm{s}}\right)\right)\right) 
\end{equation}
\vspace{-1.5em}
\begin{equation}\label{eq8}
M_g=\sigma _1\left ( C_1\left ( F_g \right )  \right ) 
\end{equation}
Among them, \emph{$F_s$} is the feature map obtained by the decoder DFB, \emph{$C_2$} means to perform a 3x3 convolution operation, \emph{$C_l$} is to use a 7×1 convolution operation, \emph{$C_v$} is to use a 1×7 convolution operation, and \scalebox{1.4}{\text{©}} is the stitching of feature maps operate.

In the gland boundary enhancement module, we perform threshold processing on the pixel values of the gland boundary to perform an enhancement operation on the boundary. The specific process is shown in equation \eqref{eq9}
\begin{equation}\label{eq9}
M_{\mathrm{b}}=\sigma_1\left(C_1\left(M_t \scalebox{1.2}{$\text{©}$} M_g\right)\right)
\end{equation}
\emph{$M_b$} is the weight map of the gland boundary, \emph{$C_l$} represents a 1×1 convolution operation, \scalebox{1.2}{\text{©}} is the splicing operation of the feature map, and \emph{$M_t$} is a threshold processing on the feature map to obtain a preliminary boundary feature map. The specific process  is shown in equation \eqref{eq10}. To better obtain the boundary features, we obtain the adaptive threshold δ from \emph{$M_g$}, which is defined as formula \eqref{eq11}. Finally, the feature enhancement operation is performed on the boundary of the gland. The specific process is shown in equation \eqref{eq12}.
\vspace{-0.5em}
\begin{equation}\label{eq10}
M_t[i, j]= \begin{cases}1 & M_g \geq \delta \\ 0 & M_g<\delta\end{cases}
\end{equation}
\vspace{-1.1em}
\begin{equation}\label{eq11}
\delta=C_1\left(P_1\left(M_{\mathrm{g}}\right)\right)
\end{equation}
\vspace{-1.1em}
\begin{equation}\label{eq12}
F_{\mathrm{s}}^{\prime}=C\left(\left(F_{\mathrm{s}} \scalebox{1.5}{$\otimes$} M_g\right) \scalebox{1.5}{$\oplus$}\left(F_{\mathrm{s}}  \scalebox{1.5}{$\otimes$} M_t\right)\right) \scalebox{1.5}{$\oplus$} F_{\mathrm{s}}
\vspace{-1em}
\end{equation}
\vspace{-1.5em}
\section{Experiments and analysis}
\label{sec:pagestyle}
\vspace{-1em}
\subsection{Datasets}
\label{ssec:subhead}
\vspace{-0.6em}
In this paper, the two data sets of the Gland Segmentation challenge (GlaS[15]) and Colorectal Adenocarcinoma Gland (CRAG[11]) are used in the experiments. The GlaS dataset has 165 H\&E-stained histopathology images extracted from 16 WSI images. The image size mostly is 775 × 522 pixels. The training set consists of 85 images, including 37 benign and 48 malignant, the test set is divided into two parts: test set A (60 images) and test set B (20 images). All experiments on GlaS are average experimental results on test sets A and B. CRAG has 213 H\&E-stained histopathology images from 38 WSI images. The scanned image size is 1512 ×1516 pixels with the corresponding instance-level ground truth. The training set has 173 images and the test set has 40 images with different cancer grades. The CRAG dataset has more irregular malignant glands, which are more difficult to identify than the GlaS dataset.
\vspace{-1.5em}
\subsection{Implementation Details and Evaluation Criteriat}
\label{ssec:subhead}
\vspace{-0.6em}
The experiments in this article used two NVIDIA GeForce GTX 3090 GPUs for training and adopted the deep learning PyTorch framework. In all experiments, we performed training for 1000 epochs, using the Adam optimizer with an initial learning rate set to 5×{$10^{-4}$} and a training batch size of 4. Due to the noise in the data set and the limited number of images, we adopted a data expansion strategy, including horizontal flipping, affine transformation, random elastic transformation, and random cropping. Enhancement operation, The size of the final image input to the network is GlaS 416×416 and CRAG 512×152. We made a triple mask for each gland and used cross-entropy loss to better supervise network learning. We used the three indicators specified by the MICCAI 2015 Challenge to evaluate the segmentation results, including F1, object-level Dice, and object-level Hausdorff distance, where F1 is to evaluate the accuracy of a single gland detection, and object-level Dice is an assessment of volume-based segmentation accuracy of individual glands, object-level Hausdorff distance evaluates the shape similarity between the segmentation result and its ground truth.
\vspace{-1.5em}
\subsection{Ablation Experiment}
\label{ssec:subhead}
\begin{figure}[t]
    \centering
    \includegraphics[width =.48\textwidth]{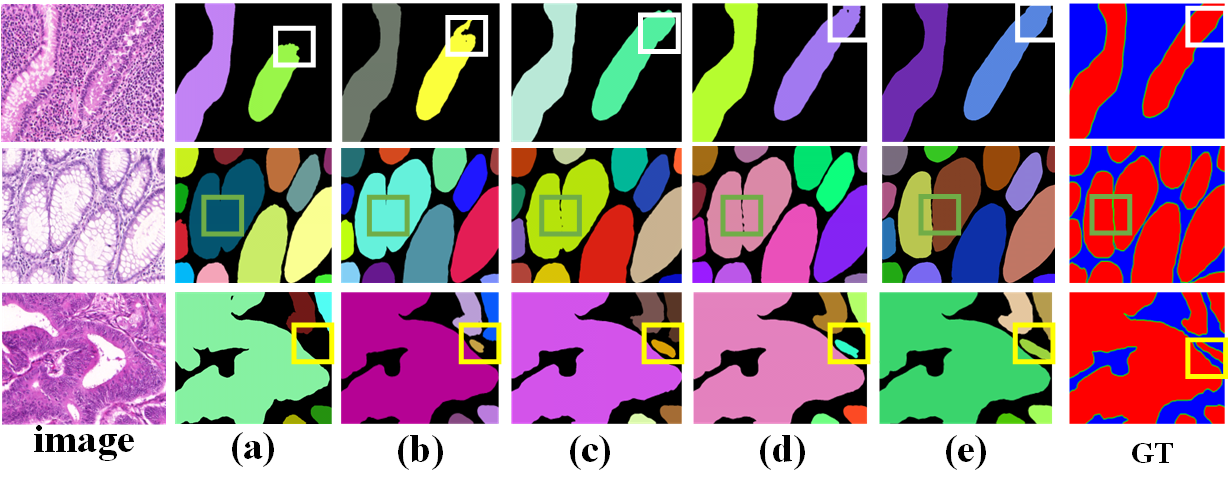}
	\vspace{-2.5em}
    \caption{(a) Backbone segmentation result, (b) Backbone+LD segmentation result, (c) Backbone+LD+FFM segmentation result, (d) Backbone+LD+FFM+DFB segmentation result, (e) DEA-Net segmentation result}
	\vspace{-1.6em}
	\label{fig4}
\end{figure}

\vspace{-0.6em}
To verify the effectiveness of our proposed module, we use U-Net as a Backbone to conduct ablation experiments on the GlaS dataset, as shown in Table \ref{table1}. The bold part represents the highest index. There are limitations in the Backbone segmentation effect. The LD module was added to Backbone, and the index was significantly improved. To better integrate features, we added the FFM module, and the index was further improved. For the gland boundary, we introduced the BEA mechanism to enhance it, and the segmentation index improved again. When we added the DFB module, F1 was 86.1\%, Dice was 89.6\%, and Hausdorff was 60.774. To better show the effectiveness of these modules, as shown in Figure \ref{fig4}, the white box in the first column and the orange box in the third column As shown in the highlighted part, the designed LD module and FFM module can solve the problem of under-segmentation and sticking of some small glands. The designed BEA mechanism and DFB module in the highlighted part of the green box in the second column can solve the problem. Difficulty in segmenting adherent glands.
\begin{table}[]
\caption{Ablation experimental results on GlaS dataset}
\centering
\label{table1}
\resizebox{.48\textwidth}{!}{
\begin{tabular}{lccc}
\hline
Methods              & F1(\%) & Dice(\%) & Huasdorff \\ \hline
Backbone            & 86.1   & 85.1     & 87.318    \\
Backbone+LD         & 87.4   & 86.9     & 72.181    \\
Backbone+LD+FFM     & 88.1   & 87.8     & 68.405    \\
Backbone+LD+FFM+BEA & 89.1   & 88.8     & 63.786    \\
DEA-Net             & \textbf{89.3}   & \textbf{89.6}     & \textbf{60.774}    \\ \hline
\end{tabular}}
\vspace{-1.5em}
\end{table}
\vspace{-1.5em}
\subsection{Performance Comparison Of Different Methods}
\label{ssec:subhead}
\vspace{-0.5em}
\begin{table}[]
\caption{Performance Comparison on Glas and CRAG}
\label{t2}
\resizebox{.45\textwidth}{!}{
\begin{tabular}{lllll}
\hline
Datasets & Methods             & F1(\%) & Dice(\%) & Hausdoff \\	\hline
         & DCAN {[}9{]}        & 81.4   & 83.9     & 102.9    \\
         & HC-FCN {[}16{]}     & 84.5   & 87.0     & 69.1     \\
         & MILD-Net {[}11{]}   & 87.9   & 87.5     & 73.7     \\
GlaS     & SADL {[}17{]}       & 88.9   & 87.3     & 76.7     \\
         & CMD-Net {[}18{]}    & 89.0   & 88.1     & 69.2     \\
         & MSFCN {[}19{]}      & 88.2   & 88.6     & 66.5     \\
         & GCSBA-Net {[}8{]}   & 87.4   & 87.5     & 72.2     \\
         & GAGLVT-Net {[}20{]} & 88.8   & 88.5     & 70.2     \\
         & DEA-Net             & \textbf{89.3}   & \textbf{89.6}     & \textbf{60.8}  \\ \cline{2-5}
         & DCAN {[}9{]}        & 73.6   & 79.4     & 218.8    \\
         & DeepLabv3           & 64.8   & 74.5     & 281.4    \\
         & MILD-Net {[}11{]}   & 82.5   & 87.5     & 160.1    \\
         & DSE {[}21{]}        & 83.5   & 88.9     & 120.1    \\
CRAG     & MSFCN {[}19{]}      & 82.5   & 89.2     & 130.4    \\
         & TA-Net {[}22{]}     & 84.2   & 89.3     & \textbf{105.2}    \\
         & 2D RNN {[}23{]}     & 82.6   & 86.5     & 127.2    \\
         & ECGSSL {[}24{]}     & 83.6   & 89.3     & 115.2    \\
         & DEA-Net             & \textbf{86.0}   & \textbf{89.9}     & {129.4}    \\ \hline
\end{tabular}}
\vspace{-1.9em}
\end{table}
To prove the effectiveness of our proposed method, this method was compared with the other eight gland segmentation models on the GlaS and CRAG datasets, and the specific results are shown in Table \ref{t2}. The bold part represents the highest index. On the GlaS data set, our proposed DEA-Net achieved the best indicators, which was 0.5\% higher than the second-place GAGLVT-Net's F1, 1.1\% higher than Dice, and 9.4 smaller Hausdoff. It can be seen the advanced nature of the method proposed by us. On the CRAG data set where malignant glands are the majority, compared with other methods, the DEA-Net we proposed reached the best level in F1 and Dice indicators and was better than the second place TA-Net The F1 increased by 1.8\%, and Dice increased by 0.6\%, which proves that our proposed method has a good generalization ability in the segmentation of malignant glands.
\vspace{-1.9em}
\section{conclusion}
\label{sec:typestyle}
\vspace{-1.1em}
To solve the problems of under-segmentation of malignant glands and difficult segmentation of tightly adherent glands in gland segmentation, this paper proposes a DEA-Net method that fuses dual encoders and boundary-enhanced attention, and guides encoding through the proposed local semantics LD and multi-scale feature fusion module FFM solve the problem of gland under-segmentation. For closely connected glands, we propose the boundary-enhanced attention mechanism BEA and the deep feature decoder DFB to better restore feature space information. Experimental results It shows that the index of our proposed method has been greatly improved compared with other gland segmentation methods. However, this article has only been tested on the gland segmentation data set. Future work can be done on different data sets. The proposed method is validated to evaluate and improve its generalization ability.


\end{CJK}
\end{document}